\documentclass[preprint]{revtex4}
\usepackage{graphicx}

\newcommand{\ket}[1] {\left| #1 \right\rangle}

\begin{document}

\title{Using machine learning to find exact analytic solutions to analytically posed physics problems}

\author{Sahel Ashhab}
\affiliation{Advanced ICT Institute, National Institute of Information and Communications Technology, 4-2-1, Nukuikitamachi, Koganei, Tokyo 184-8795, Japan}

\begin{abstract}
We investigate the use of machine learning for solving analytic problems in theoretical physics. In particular, symbolic regression (SR) is making rapid progress in recent years as a tool to fit data using functions whose overall form is not known in advance. Assuming that we have a mathematical problem that is posed analytically, e.g.~through equations, but allows easy numerical evaluation of the solution for any given set of input variable values, one can generate data numerically and then use SR to identify the closed-form function that describes the data, assuming that such a function exists. In addition to providing a concise way to represent the solution of the problem, such an obtained function can play a key role in providing insight and allow us to find an intuitive explanation for the studied phenomenon. We use a state-of-the-art SR package to demonstrate how an exact solution can be found and make an attempt at solving an unsolved physics problem. We use the Landau-Zener problem and a few of its generalizations as examples to motivate our approach and illustrate how the calculations become increasingly complicated with increasing problem difficulty. Our results highlight the capabilities and limitations of the presently available SR packages, and they point to possible modifications of these packages to make them better suited for the purpose of finding exact solutions as opposed to good approximations. Our results also demonstrate the potential for machine learning to tackle analytically posed problems in theoretical physics.
\end{abstract}

\maketitle

\section{Introduction}
\label{Sec:Introduction}

Machine learning has found applications in a large number of areas, especially in the area of science and technology. One of the most remarkable achievements that highlight the power of machine learning is that, given only the basic rules of chess, a machine was able to train itself and subsequently defeat any present-day competitor, human or computer running older methods \cite{Silver}. Machines are also now able to diagnose diseases with a prediction accuracy that is comparable to that of well-trained and experienced physicians \cite{NatMatEditorial}.

There are now a vast number of proposals to use machine learning for scientific research in such a way that machines perform tasks that were conventionally performed by researchers \cite{Radovic,Carleo,Iten,Qiu,Alhousseini}. For example, when scientists deal with experiments that generate large amounts of data and they look for specific features in the data, it is natural to use automated methods to look for these features. In conventional machine learning applications, the overall structure of the data is known, and the algorithms determine the parameters that give the best fit for the data within the known structure. More recent algorithms, such as deep learning, allow one to make predictions even in cases when the rules of inference are unknown, e.g.~when it is not known exactly what features one needs to look for. The algorithm is given a training data set, and it finds patterns in this data that it then uses to make inference about new data, e.g.~the so-called test data set.

An active and still developing area of machine learning is symbolic regression (SR)  \cite{Koza,Schoenauer,Schmidt,Eureqa,Gaucel,Wu,Jovanovic,Hernandez,Guimera,Udrescu,Udrescu2,Bomarito,Keren,Fajardo,Cornelio,Makke,Kubalik2023,Oh,Cao,CoryWright}. In scientific fitting problems, and taking the single-variable case for simplicity, one is typically given a set of $N$ data points of the form $(x_j,y_j)$, with $j=1,2,...,N$, and one is interested in finding a function $f(x)$ such that one can say that, to some approximation, $y=f(x)$. In the conventional and common case, the general form of $f(x)$ is known, and only some fitting parameters need to be determined. For example, if one is looking for a linear fit, the function $f(x)$ can be expressed as $f(x)=\alpha x + \beta$, where $\alpha$ and $\beta$ are the fitting parameters. In the study of oscillatory dynamics, one might use the function $f(x)=\alpha \sin (\omega x+\phi)$, where $\alpha$, $\omega$ and $\phi$ are fitting parameters. In SR, even the form of $f(x)$ is unknown at the beginning of the fitting process. For example, one might not know in advance whether the best fit to the data will be a polynomial function, a trigonometric function, an exponential function or some combination (including sums, products and concatenations) of these. For example, the best fit could be a function of the form
\begin{equation}
f(x) = \alpha + \beta x + \gamma x^2+ \delta e^{\eta x^2 + \kappa x} \sin (\omega x + \phi),
\end{equation}
keeping in mind that even the general form of the function is not known before finding that it gives the best fit. We note here that the generalization to problems with more than one variable, i.e.~replacing $x$ by multiple input variables, is conceptually straightforward although it can make the practical implementation of the algorithm to solve a given problem significantly more time consuming.

Previous studies on SR \cite{Koza,Schoenauer,Schmidt,Eureqa,Gaucel,Wu,Jovanovic,Hernandez,Guimera,Udrescu,Udrescu2,Bomarito,Keren,Fajardo,Cornelio,Makke,Kubalik2023,Oh,Cao,CoryWright}, which have mainly been focused on developing the technique, have generally taken the approach of assuming that the input is given in the form of data and we would like to find the best fit for the data. A number of these studies explored the potential of SR to solve physics problems. For example, Schoenauer {\it et al.} \cite{Schoenauer} used SR to discover phenomenological models to describe the mechanical properties of new materials based on experimental data. Schmidt and Lipson \cite{Schmidt} used SR to identify conservation laws in experimental motion-tracking data of a few different mechanical systems. Gaucel {\it et al.} \cite{Gaucel} applied SR to discover differential equations from data. Hernandez {\it et al.} \cite{Hernandez} developed an algorithm to derive simple models for the inter-atomic potentials from {\it ab-initio} calculation data. Udrescu and Tegmark \cite{Udrescu,Udrescu2} developed an algorithm that incorporates a variety of common physics problem solving techniques, such as inspecting units and symmetries in the data. Bomarito {\it et al.} \cite{Bomarito} used SR to find simple mathematical models for the plasticity of materials based on microscopic simulation data. Kubal\'{i}k {\it et al.} \cite{Kubalik2023} developed an algorithm in which constraints are included via additional cost functions whose values are minimized if the constraints are satisfied. Oh {\it et al.} \cite{Oh} developed an algorithm that searches for a function that simultaneously fits the available data and satisfies a differential equation that the data is expected to obey. Testing whether the function satisfies the differential equation is performed via automated symbolic differentiation and algebraic manipulation. Cao {\it et al.} \cite{Cao} built on the work of Oh {\it et al.} \cite{Oh} and added a pruning step in which the algorithm looks for redundancies and, if possible, simplifies the appearance of the fitting function. It should be noted that some of these studies, e.g.~Refs.~\cite{Gaucel,Hernandez,Udrescu,Oh,Cao}, tackled mathematical and physical problems with known solutions and in some cases generated the data from the known solution, and the authors found these known solutions using SR. It is important to note, however, that the algorithms used in these studies were designed to simply find a good fit to the data and attempt to minimize a fitting error metric. In fact, in most cases, the desired function was obtained in spite of the fact that the researchers added noise to the data in order to simulate realistic experimental situations.

Here we take a different approach: our aim is to use SR as a tool to find exact analytic solutions to mathematical problems that are themselves expressed analytically. In other words, we deal with conventional theoretical physics problems. There is no numerical data in the input or in the output. The problem can be a set of equations. The sought solution is also expected to be an equation or set of equations. Numerical data is introduced only as a tool in the intermediate step of trying to solve the equations. The idea of our approach is perhaps best explained by the example of the Landau-Zener problem, which we describe in the following.

{\bf Landau-Zener problem}

Here we describe a physical problem that illustrates the possible usefulness of SR in solving analytically formulated problems. In 1932, four physicists [Landau, Zener, St\"uckelberg and Majorana (LZSM)] independently addressed the same problem that is commonly known as the Landau-Zener (LZ) problem \cite{DiGiacomo,Shevchenko}. The problem can be formulated as follows: we have a time-dependent quantity (the quantum state vector) composed of two complex numbers $\psi(t) = (\psi_{\uparrow}(t), \psi_{\downarrow}(t))$ that obeys the Schr\"odinger equation
\begin{equation}
\left( \begin{array}{c} i\frac{d\psi_{\uparrow}}{dt} \\ i\frac{d\psi_{\downarrow}}{dt} \end{array} \right) = \frac{1}{2}
\left( \begin{array}{cc} v t & \Delta \\ \Delta & - v t \end{array} \right) \cdot \left( \begin{array}{c} \psi_{\uparrow} \\ \psi_{\downarrow} \end{array} \right),
\label{Eq:TwoLevelLZSE}
\end{equation}
where $v$ and $\Delta$ are parameters that depend on the exact conditions of the problem (e.g.~the rate of change and the minimum value of a magnetic field applied to the magnetic dipole of an electron). We have not explicitly written ``$(t)$'' in the above equation, but it should be kept in mind that $\psi_{\uparrow}$ and $\psi_{\downarrow}$ are functions of the time variable $t$. We can, with no loss of generality, assume that $v$ and $\Delta$ are both positive. It is assumed that at the initial time $t\rightarrow -\infty$ the quantum state is given by $(\psi_{\uparrow}, \psi_{\downarrow}) = (1,0)$. We now want to know the values of $|\psi_{\uparrow}|^2$ and $|\psi_{\downarrow}|^2$ at the final time $t\rightarrow \infty$. These quantities represent the probabilities that the quantum system will stay in its initial state or make a so-called LZ transition. If we make the hypothetical assumption that the solution to this problem is not well known, solving the problem is not at all straightforward. In 1932, LZSM solved the problem using special functions and other mathematical methods that are not familiar to most physicists. In fact, Landau did not rigorously solve the problem; he only solved it in two opposite limits and then made an intuitive guess for the general solution.

Using dimensional analysis, or alternatively by inspection of Eq.~(\ref{Eq:TwoLevelLZSE}), it is relatively easy to recognize that the solution, i.e.~$|\psi_{\uparrow}(t\rightarrow\infty)|^2$ or $|\psi_{\downarrow}(t\rightarrow\infty)|^2$, must be a function of the single parameter $\Delta^2/v$. It is not easy to make much progress beyond this point. Although the LZ problem is difficult to solve, its solution is remarkably simple:
\begin{equation}
|\psi_{\uparrow}(t\rightarrow\infty)|^2 = e^{-\frac{\pi\Delta^2}{2v}},
\end{equation}
which is the solution obtained by LZSM.

This situation leads to the following idea: if (hypothetically) we wanted to solve the LZ problem but we did not know the above simple solution, we could numerically solve the Schr\"odinger equation and find $|\psi_{\uparrow}(t\rightarrow\infty)|^2$ for various values of $\Delta^2/v$. Each one of these data points can be generated with an accuracy of several significant figures in a fraction of a second on a present-day personal computer. Simply by plotting the data of $|\psi_{\uparrow}(t\rightarrow\infty)|^2$ as a function of $\Delta^2/v$ and looking at the resulting graph, one would immediately recognize that the plot looks like an exponential function. A straightforward fitting procedure would yield the function $f(x) = e^{-\pi x/2}$ and show that the deviation between the data and fitting function is at the level of numerical errors in the computation, indicating that this function is in fact the exact solution and not simply a convenient approximation. We emphasize here that in this example, the numerical data and fitting procedure are used only in the intermediate steps. The goal is finding an exact analytic solution. We also emphasize that in theoretical physics having an exact solution is fundamentally different from having good approximations that allow us to make reliable numerical predictions. For example, even though we had a set of good approximations for the quantum Rabi model (QRM) covering essentially all parameter regimes \cite{Ashhab2010}, the discovery of integrability and exact solution was an important achievement in the study of the QRM \cite{Braak}. Exact solutions often help elucidate the physical mechanisms at play in the phenomenon under study, and they can allow generalizing known results to new regimes.

For the LZ problem, knowledge about basic mathematical functions is sufficient to determine the function that fits the data. The question now is: if we are dealing with a problem that yields data that is not easily recognizable to a human scientist inspecting the data, can a machine-learning algorithm identify the function? For example, for functions that take more than one input variable [e.g.~$f(x, y, z)$ which is a function of three variables with nontrivial functional behavior], it would in general be difficult for a human to discern all the different features and trends in the function from the data. This function identification procedure is exactly what SR does.

The important point to note here is that there is a class of theoretical physics (and applied mathematics) problems that have no known analytic solutions but for which the solution can be evaluated straightforwardly on a computer for any value of the input parameter (or parameters). Furthermore, exactly solvable theoretical physics problems, even ones that can be difficult to solve analytically, often have simple-looking solutions that contain only a few terms and only a few of a standard set of mathematical functions. It is for these problems that our approach would be most relevant and effective. Needless to say, having an analytic closed-form solution is far more attractive and valuable than just knowing that the computer can reliably give us the value of the function for any set of input parameters.

To summarize the solution strategy, one starts with a problem posed as a set of equations. One then generates numerical data for a set of input variable values. One then uses SR to fit the data and hopes that the fitting function generated by SR is the exact solution to the problem.

It should be noted here that some studies on complicated physical systems, e.g.~density functional theory and molecular dynamics, also generate data and then use general function fitting techniques to obtain models for atom-atom interactions \cite{Unke,Mishin,Behler}. However, in these studies, it is assumed and accepted from the beginning that the obtained models will be approximate empirical models, not exact solutions. In this context, the concept of computational irreducibility \cite{Wolfram,Rowland} is relevant: some problems can be transformed to problems whose solutions are either known or easily obtainable, while other problems do not allow any simplification without approximations. Many-body physics problems are almost always computationally irreducible. As we shall discuss below, the fact that we assume the existence of analytic exact solutions and that we seek to find these solutions leads to some constraints that we can impose on the computation. It also leads to other differences in applying SR to the problems that we try to solve.

We emphasize that in this work we do not develop a new SR algorithm. We use a state-of-the-art SR package \cite{Udrescu,Udrescu2} to tackle a few physics problems and search for their exact solutions. These case studies allow us to assess the ability of currently available SR packages to find the sought solutions. They also allow us to identify some possible future modifications that can make SR algorithms better suited for the purpose of finding exact solutions.

\section{Method}
\label{Sec:Method}

In this section we discuss various aspects of implementing the approach described in Sec.~\ref{Sec:Introduction}.

\subsection{Algorithm structure}

The procedure for solving the problem can go as follows:

(1) The user chooses a number input variable values to cover a good portion of the variable space. The points could be spaced equally or unequally, e.g.~randomly.

(2) The user numerically evaluates the values of the output variable for the selected points.

(3) The user applies the SR algorithm to find a mathematical function that describes the relation between the input and output variables.

If the above procedure does not produce the exact solution being sought, which could happen for hard problems, the following steps can be added:

(4) After starting with the initial set of input variable values, if several possible candidates for the sought function all meet the required fitting criteria (which is expected to be a rare possibility when seeking an exact solution), the user can start to actively choose values of $x$ to help distinguish between the different candidate choices for $f(x)$ and eliminate incorrect ones until the correct function is identified with a high degree of confidence. The locations of the new data points can be decided so as to maximize the distinguishability between the different candidate solutions. Needless to say, the choice of the input variable values for the new data points can be automated, with the locations of the new points decided based on where the different candidate solutions have maximum separation, and the human user does not need to make the decision about where to place these new points. Simply put, our approach can incorporate ideas similar to active learning \cite{Settles}.

(5) The user can focus on specific limits, i.e.~input variable values close to specific points, to obtain the functional form of the solution in these limits. Knowing the behavior of the function in different limits can be helpful for the purpose of inferring the general form of the function, just like how Landau was able to solve the LZ problem.

(6) It is common in physics problems that the problem simplifies in a variety of limits, such that various asymptotic values of the solution can be inferred either from solving very simple equations or even based on intuitive arguments, without solving the equations in detail. In some cases, the solution might be known for specific values of the input variables, e.g.~when the input variables are equal to zero. Furthermore, there can be physicality conditions on the allowed range of the solution, e.g.~probabilities being bounded between zero and one. Any such prior knowledge about the solution can be incorporated as constraints to be used during the data fitting.

(7) In the case of multiple-variable functions, the user can fix the values of some input variables and try to find the output variable as a function of the remaining input variables in various special cases. The special-case solutions can then serve as guides or be used as constraints in the next step of searching for the full solution.

(8) Zeros, as well as extrema, can be used to assist the search for fitting functions. Different mathematical functions have their own characteristic patterns of where the zeros are located. The zeros in the data can therefore serve as a sort of fingerprint for the sought function.

We should note that some of the points listed above overlap with similar points discussed in the SR literature, e.g.~\cite{Udrescu,Udrescu2,Kubalik2023}.

\subsection{Noise in the data}

One point that is worth noting here is that data fitting algorithms usually assume that we are dealing with data that has some noise component, e.g.~because of experimental errors or because many unknown factors contribute to determining the data values. As a result, one searches for the best fit that gives the lowest net deviation from the data (among the set of candidate fitting functions). Sometimes, outliers are discarded based on the assumption that they are not representative of the bulk of the data. In the situation that we are considering in this work, we generally assume that the value of the function for any set of input variables can be calculated numerically using well-established numerical methods with known uncertainty levels. It is therefore possible, at least for some problems, to keep numerical errors at arbitrarily small levels for all data points. In other words, unlike many problems that are commonly encountered in the field of machine learning, one can assume that there is no noise or random component in the data (up to the small numerical errors in the computational evaluation of the function), no outliers that can be ignored, no missing data etc. Therefore one does not need to design the algorithm to accommodate some deviation from the data. One can require that these deviations remain at the error level in the computations, say $10^{-10}$. This fact can be incorporated into the algorithm via the cost function that penalizes deviations between the fitting function and the data. One can use a function that remains negligibly small up to the known error level of the data and then increases dramatically above the error level. In other words, one could disqualify a candidate solution because it gives a deviation that is higher than the allowed value for a single data point, even if this candidate solution gives the best fit for most of the data points. This procedure could be helpful in eliminating disqualified solutions and avoiding the common problem in which the search algorithm gets trapped in a local minimum in the landscape of the cost function.

In relation to the question of noise in the data, it is worth mentioning the Bayesian machine scientist algorithm developed in Ref.~\cite{Guimera} and the abrupt transition between a low-noise regime and a high-noise regime observed in that context \cite{Fajardo}. In the case of low noise, the algorithm is well suited to identify the correct solution as the unique best solution. In the case of sufficiently high noise, multiple solutions have comparable fitting quality metrics and are therefore similarly plausible candidates for the sought solution. Some incorrect solutions can even acquire an advantage by virtue of their low complexity, considering that low complexity is commonly rewarded in SR algorithms. As a result, it can become practically impossible for the algorithm to distinguish between multiple candidates and/or identify the correct solution.

\subsection{Running time}

The success of the algorithm obviously relies on the existence of the exact solution in the set of candidate solutions considered by the algorithm. The time needed to perform a full calculation will scale with the number of functions in the full set of candidate functions. One can then start with a relatively small set, e.g.~with a limited set of building blocks and a limited number of the building blocks in the candidate solutions. If the first attempt with the small set of candidate solution fails, one can gradually increase the size of the set. The algorithm can still fail for any one of multiple reasons, for example (1) the nonexistence of an exact solution within the set of standard functions considered by the algorithm or (2) the high complexity of the exact solution that would lead to a prohibitively long computation time.

\subsection{Verification}

One question can be how to verify that an obtained solution is indeed the exact solution. In some cases, finding a fitting function that does not deviate from any data point by more than the numerical error level is a strong indication that the solution is the exact solution. For example, if we are dealing with a problem that has only one or a few input variables, the data points are chosen to cover all parameter regimes, and the deviation between the data and fitting function remains below a numerical error level of, say, $10^{-10}$, we can have extreme confidence that we have indeed found the exact solution. In other cases, e.g.~if we are trying to find the full solution to a differential equation, checking whether a given function is the solution is sometimes a very simple task: one determines the relevant derivatives, substitutes these derivatives in the differential equation and checks if the equation is satisfied.

It should be noted here that, strictly speaking, fitting the data to within the numerical error level is a necessary but not sufficient condition for an exact solution. One could imagine a problem for which there is an approximate solution that is accidentally very close to the exact solution. In this context, it is interesting to note the occasional existence of accidentally excellent approximations, such as 1 year $\approx\pi\times 10^7$ seconds, which is accurate to within a relative error of 0.004. While we cannot completely eliminate the possibility that a similar situation could occur when validating a candidate solution to a mathematical problem, it seems far less likely for this situation to occur when dealing with a function over a range of input values than it is in the case of having an accidentally good approximation for a single number. Furthermore, if a candidate solution exhibits full agreement with the data within the numerical error level, it is possible to apply extra scrutiny by generating additional numerical data with a lower error level to confirm that the candidate solution still fully agrees with the data to within the reduced error level. Nevertheless, from the point of view of mathematical rigor, the only way to be sure that a candidate solution is the exact solution is with analytic verification. As mentioned above, this situation can indeed occur, for example when testing possible solutions of differential equations, since differentiation and algebraic manipulation of functions are typically much easier than solving differential equations.

\subsection{High-dimensional problems}

In typical machine learning applications, the most challenging problems are high-dimensional ones with a large number of input variables. The performance of machine learning approaches generally degrades for such problems. The degradation for SR might be especially rapid, considering that the increasing number of variables translates to more functional combinations that can be constructed. With this point in mind, we expect that our approach is most effective for low-dimensional problems. It should be mentioned here that high-dimensional physics problems are less likely to allow exact solutions anyway. An interesting possibility in this context is that a high-dimensional problem might lead to an exact relation between a small number of variables, in a manner similar to conservation laws. It will be interesting if such relations are discovered using SR.

\section{Results and discussion}
\label{Sec:TestAIF}

In this section we report on attempts that we made using the state-of-the-art SR package AI Feynman (AIF) to solve physics problems following the approach described above. In all of the tests we used the settings BF\_try\_time=60, BF\_ops\_file\_type= ``14ops.txt'', polyfit\_deg=3 and NN\_epochs=500. We tried different settings in some of the calculations, especially those in which AIF did not produce the correct solutions with the basic settings. However, the different settings that we tried did not lead to improved solutions. It should also be noted that AIF does not necessarily generate the same suggested solutions on repeated runs of the same problem. We did 10 runs for each problem. In cases where we list several suggested solutions, we report the results that were obtained on the first run. In these cases, no clear improvement in the best suggested solution was obtained in the subsequent runs. In cases where we give a single solution obtained by AIF, we take the best solution obtained from the all the runs. The best solution was always found in the first few runs. At some point, the additional runs simply reproduced suggested solutions that can be found in the first few runs.

\subsection{Solved problems}
\label{Sec:SolvedProblems}

As a first reference point, we took the LZ probability function $P=e^{-\pi/(2x)}$ with $x=v/\Delta^2$, and we used this function to generate 121 data points with $x$ distributed uniformly on a logarithmic scale in the range $[10^{-3},10^3]$. The data was generated using the software Wolfram Mathematica as well as using the numpy package in python. The maximum difference between the $P$ values in the two data sets was about $1.1 \times 10^{-16}$. We can therefore estimate the error in both data sets to be at or below the level of $10^{-16}$. We ran AIF to find the best fit for the data.  The solution file created by AIF contained the following six suggested solutions, ordered from the solution with the smallest error at the top to the solution with the largest error (but smallest complexity) at the bottom
\begin{eqnarray}
f_1(x) & = & ( \exp (-3.141592653589793/x) )^{0.5}
\nonumber \\
f_2(x) & = & 0.000000000000 + \sqrt{\exp ( (\pi+\sin\pi) / (-x) )}
\nonumber \\
f_3(x) & = & (\exp ( -3 / x) )^{0.5}
\nonumber \\
f_4(x) & = & \tan ( 2 * \exp ( - \exp (1/x) ) )
\nonumber \\
f_5(x) & = & 3 * \exp ( - \exp (1/x) )
\nonumber \\
f_6(x) & = & 0.
\label{Eq:AIFSolutionsLZ}
\end{eqnarray}
The first two of these solutions, i.e.~$f_1(x)$ and $f_2(x)$, are clearly the correct solutions (noting here that the long constant in $f_1(x)$ is $\pi$ to all shown significant figures). In other words, AIF succeeded in finding the exact solution to this problem. It is interesting that $f_2(x)$ contains two clearly superfluous terms: the zero term at the beginning and the term $\sin\pi$. It is also not clear why $f_1(x)$ is assigned the largest complexity value by AIF, although it looks simpler than some of the other solutions, e.g.~$f_2(x)$.

Now we move somewhat closer to the scenario described in previous sections. We generate the data using numerical simulations of the time-dependent Schr\"odinger equation. The problem that we seek to solve is similar to the LZSM problem, but with a two-level system (TLS) coupled to a harmonic oscillator \cite{Wubs}. The Schr\"odinger equation is expressed as
\begin{equation}
i\frac{d\ket{\psi}}{dt} = \hat{H} \ket{\psi}
\end{equation}
with an infinite-dimensional complex vector $\ket{\psi}$, the Hamiltonian matrix
\begin{equation}
\hat{H} = - \frac{\Delta}{2} \hat{\sigma}_x \otimes \hat{1}_{\infty} - \frac{vt}{2} \hat{\sigma}_z \otimes \hat{1}_{\infty} + \omega \hat{1}_{2} \otimes \hat{a}^{\dagger} \hat{a} + g \hat{\sigma}_z \otimes \left( \hat{a} + \hat{a}^{\dagger} \right),
\label{Eq:LZwOscillatorHam}
\end{equation}
$\hat{\sigma}_{\alpha}$ (with $\alpha=x,y$ or $z$) are the two-dimensional Pauli matrices, $\hat{a}$ and $\hat{a}^{\dagger}$ are, respectively, the harmonic oscillator annihilation and creation operators, $\hat{1}_{n}$ is the $n$-dimensional identity matrix, and $\otimes$ stands for the tensor product. We assume that the initial state at $t\rightarrow -\infty$ is the ground state of the Hamiltonian $\ket{\downarrow,0}$ [where the symbol $\downarrow$ indicates the TLS eigenvector with $\hat{\sigma}_z$ value $-1$, and the index 0 indicates the harmonic oscillator eigenvector with 0 photons [keeping in mind that the interaction with the TLS modifies the photon number operator from the usual $\hat{a}^{\dagger} \hat{a}$ to $(\hat{a}^{\dagger}-g/\omega) (\hat{a}-g/\omega)$ for the TLS state $\downarrow$ and to $(\hat{a}^{\dagger}+g/\omega) (\hat{a}+g/\omega)$ for the TLS state $\uparrow$]. For the initial state $\ket{\downarrow,0}$, no states of the form $\ket{\downarrow,n}$ with $n\geq 1$ will be occupied at the final time $t\rightarrow\infty$. Furthermore, if we consider a theoretical model where the states $\ket{\downarrow,n}$ with $n\geq 1$ in the TLS-oscillator system are ignored, there exists an exact solution for the final-time occupation probabilities \cite{Demkov,Ashhab2023}. The probabilities are given by
\begin{eqnarray}
P(\downarrow, 0) & = & e^{-\pi\Delta^2/(2v)}
\nonumber \\
P(\downarrow, n) & = & 0 \hspace{2.5cm} {\rm for \ all} \ n\geq 1
\nonumber \\
P(\uparrow, 0) & = & 1 - e^{-\pi\Delta_0^2/(2v)}
\nonumber \\
P(\uparrow, 1) & = & e^{-\pi\Delta_0^2/(2v)} \left( 1 - e^{-\pi\Delta_1^2/(2v)} \right)
\nonumber \\
P(\uparrow, 2) & = & e^{-\pi\Delta_0^2/(2v)} e^{-\pi\Delta_1^2/(2v)} \left( 1 - e^{-\pi\Delta_2^2/(2v)} \right)
\nonumber \\
& \vdots &
\label{Eq:LZwOProbabilities}
\end{eqnarray}
with
\begin{equation}
\Delta_n = \frac{1}{\sqrt{n!}} \left( \frac{-2g}{\omega} \right)^n e^{-2(g/\omega)^2} \Delta.
\end{equation}
This problem provides a good test case, as the complexity of the solutions increases gradually with increasing $n$, and the exact solutions contain only simple mathematical functions.

To generate the data points, we solved the Schr\"odinger equation numerically following the approach used in Ref.~\cite{Ashhab2016}: we first fixed the value of $g/\Delta$. As a specific example, we took $g/\Delta=0.1$. A polaron transformation was applied, such that we only need to keep the (polaron-transformed) states $\ket{\downarrow, 0}, \ket{\uparrow, 0}, \ket{\uparrow, 1}, \ket{\uparrow, 2}, ...$ in the simulation. We kept 101 states, going up to the state $\ket{\uparrow, 99}$. Although the initial and final times are infinite in the theoretical formulation of the problem, we set finite time boundary conditions. To determine suitable initial and final time values, we note that the dynamics is characterized by a series of probability mixing events at which sudden transitions and probability rearrangement occur. We determined the $t$ values at which these events occur, which is easy to do based on the energy-level crossing points for $g=0$. We then set the initial value of $t$ to be $100\Delta/v$ before the first event and the final value of $t$ to be $100\Delta/v$ after the last event. The total time duration was divided into $10^5$ time steps. The Hamiltonian was approximated as being constant during each time step, with its value calculated based on the $t$ value at the middle of the time step. We used 101 values of $v/\Delta^2$ ranging from $10^{-2}$ to $10^2$ and distributed uniformly on a logarithmic scale. Generating a single data point takes a few minutes on a single core of a present-day computer. Generating many data points with different values of $v/\Delta^2$ does not add much to the computation time, because all the simulations for different values of $v/\Delta^2$ require diagonalizing the same set of Hamiltonian matrices, such that some intermediate steps in the computation are performed once and do not need to be repeated for each value of $v/\Delta^2$. It is almost certain that this computation can be optimized further to generate the data faster. However, since the simulation timescale is not a limiting factor to our overall calculation, we do not attempt to find the most optimal implementation of the dynamics simulation. The $P(\downarrow, 0)$, $P(\uparrow, 0)$, $P(\uparrow, 1)$ and $P(\uparrow, 2)$ values obtained from these simulations had respective deviations of up to $1\times 10^{-5}$, $1\times 10^{-5}$, $1\times 10^{-6}$ and $9\times 10^{-9}$ from the exact formulae given in Eq.~(\ref{Eq:LZwOProbabilities}). We emphasize that these numerical error levels can be reduced by orders of magnitude if we use a larger time range and smaller time steps, albeit at the cost of a longer computation time.

\begin{figure}[h]
\includegraphics[width=10.0cm]{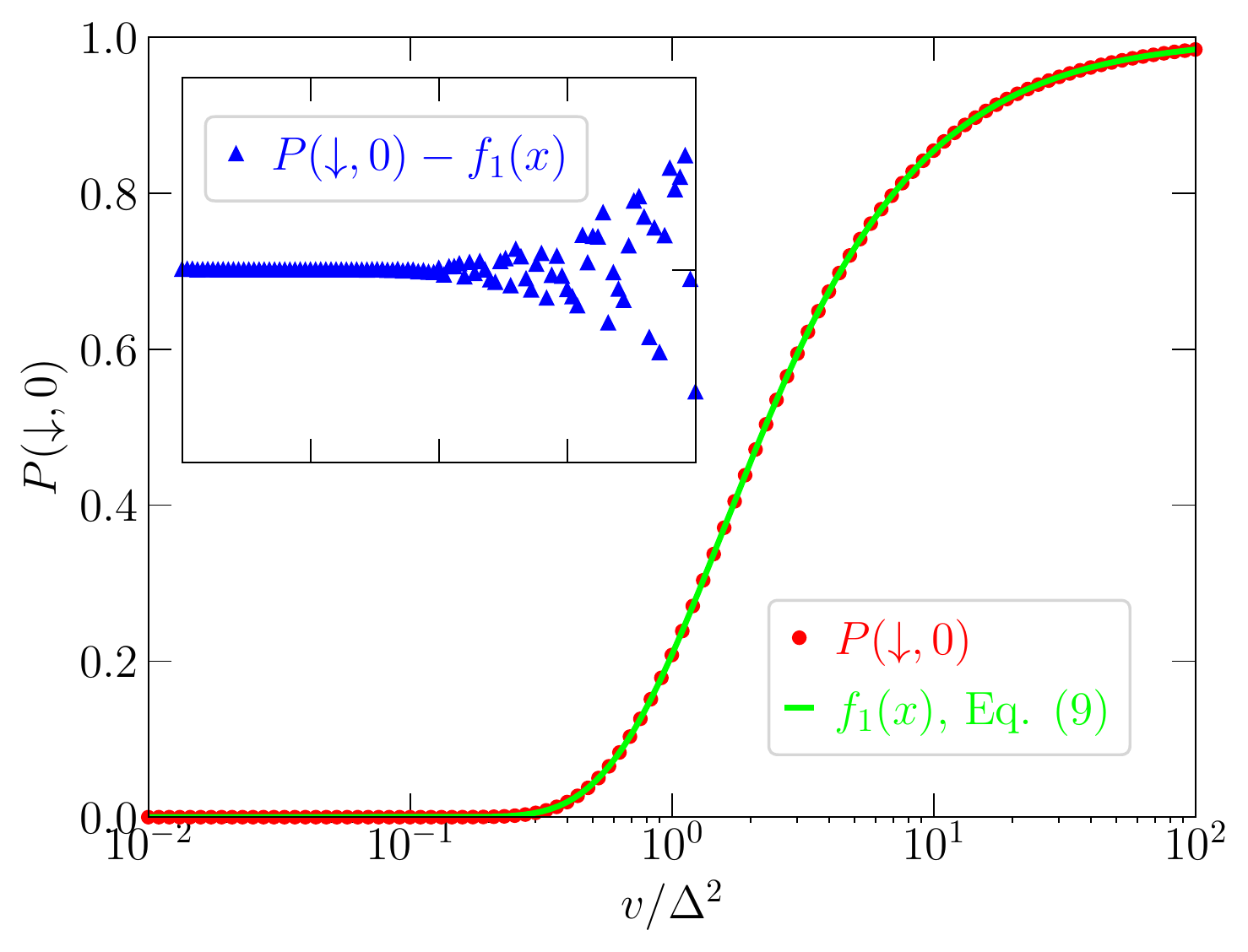}
\caption{Probability $P(\downarrow, 0)$ as a function of $v/\Delta^2$ for the LZ problem of a TLS coupled to a harmonic oscillator with $g/\omega=0.1$. The red dots are the data points that were generated by numerically solving the Schr\"odinger equation. The green line corresponds to the fitting function $f_1(x)$ in Eq.~(\ref{Eq:AIFSolutionsLZwO01E}) with $x=v/\Delta^2$. The inset shows the deviation between the red dots and green line in the main plot. The $x$ axis range in the inset is the same as that in the main plot, while the $y$ axis range in the inset is $[-2 \times 10^{-5}, 2 \times 10^{-5}]$. The deviation does not exceed the numerical error level in the data, hence meeting the criterion required for a fitting function to be an acceptable candidate for the exact solution.}
\label{Fig:LZwO01E}
\end{figure}

We ran AIF to find the best fit for the four data sets. For $P(\downarrow, 0)$, and setting $x=v/\Delta^2$, AIF gave the following suggested solutions
\begin{eqnarray}
f_1(x) & = & \exp (-1.570796326794897/x)
\nonumber \\
f_2(x) & = & 0.000000002179 + \sqrt{\exp (\pi / (-x))}
\nonumber \\
f_3(x) & = & 0.000000002179 + \exp (\pi / (x \times (\cos\pi -1)))
\nonumber \\
f_4(x) & = & \exp (-1.5 / x)
\nonumber \\
f_5(x) & = & \exp (-2 / x)
\nonumber \\
f_6(x) & = & 0.
\label{Eq:AIFSolutionsLZwO01E}
\end{eqnarray}
The constant in $f_1(x)$ is $\pi/2$ to all shown significant figures. The data and the fitting function $f_1(x)$ are shown in Fig.~\ref{Fig:LZwO01E}. The deviation between the data and $f_1(x)$ does not exceed the numerical errors from solving the Schr\"odinger equation to generate the data, as is expected for the exact solution. Apart from the first term in $f_2(x)$ and $f_3(x)$, i.e.~$2.179\times 10^{-9}$, the first three functions are all correct, only expressed in different ways.

\begin{figure}[h]
\includegraphics[width=10.0cm]{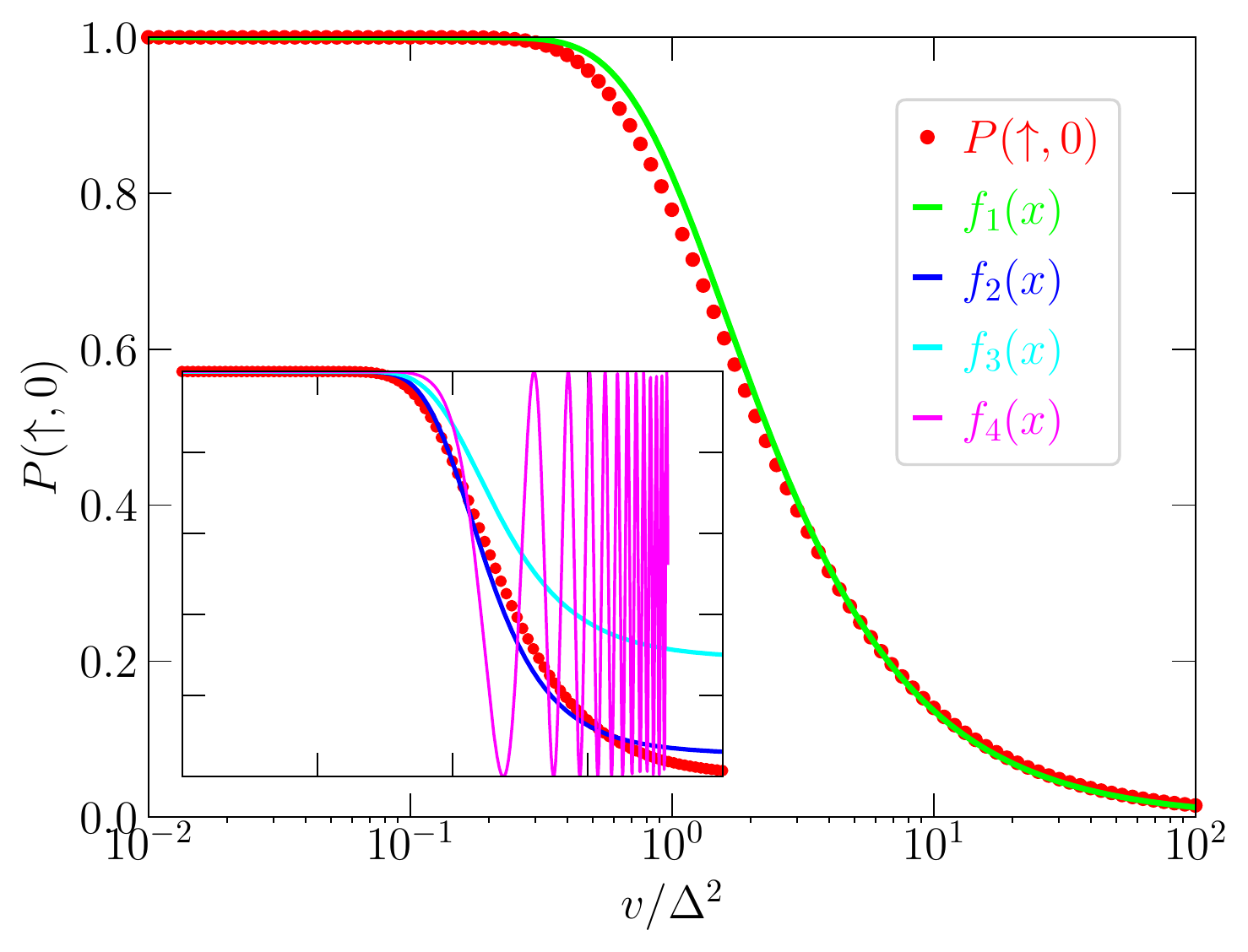}
\caption{Probability $P(\uparrow, 0)$ as a function of $v/\Delta^2$ for the LZ problem of a TLS coupled to a harmonic oscillator with $g/\omega=0.1$. The red dots are the data points that were generated by numerically solving the Schr\"odinger equation. The green, blue, cyan and magenta lines correspond, respectively, to the fitting functions $f_1(x)$, $f_2(x)$, $f_3(x)$ and $f_4(x)$ in Eq.~(\ref{Eq:AIFSolutionsLZwO010}) with $x=v/\Delta^2$. The functions $f_2(x)$, $f_3(x)$ and $f_4(x)$ are plotted in the inset to avoid crowding the main plot, which is intended to highlight the best solution generated by AIF, i.e.~$f_1(x)$. The axis ranges in the inset are the same as those in the main plot. Although it has the wrong functional form, $f_1(x)$ is a good fit to the data almost everywhere, except around $x=1$, where it deviates sufficiently for us to conclude that it cannot be the exact solution. The magenta line in the inset is plotted only up to $v/\Delta^2=40$ to avoid obscuring the blue and cyan lines at large values of $x$.}
\label{Fig:LZwO010}
\end{figure}

\begin{figure}[h]
\includegraphics[width=10.0cm]{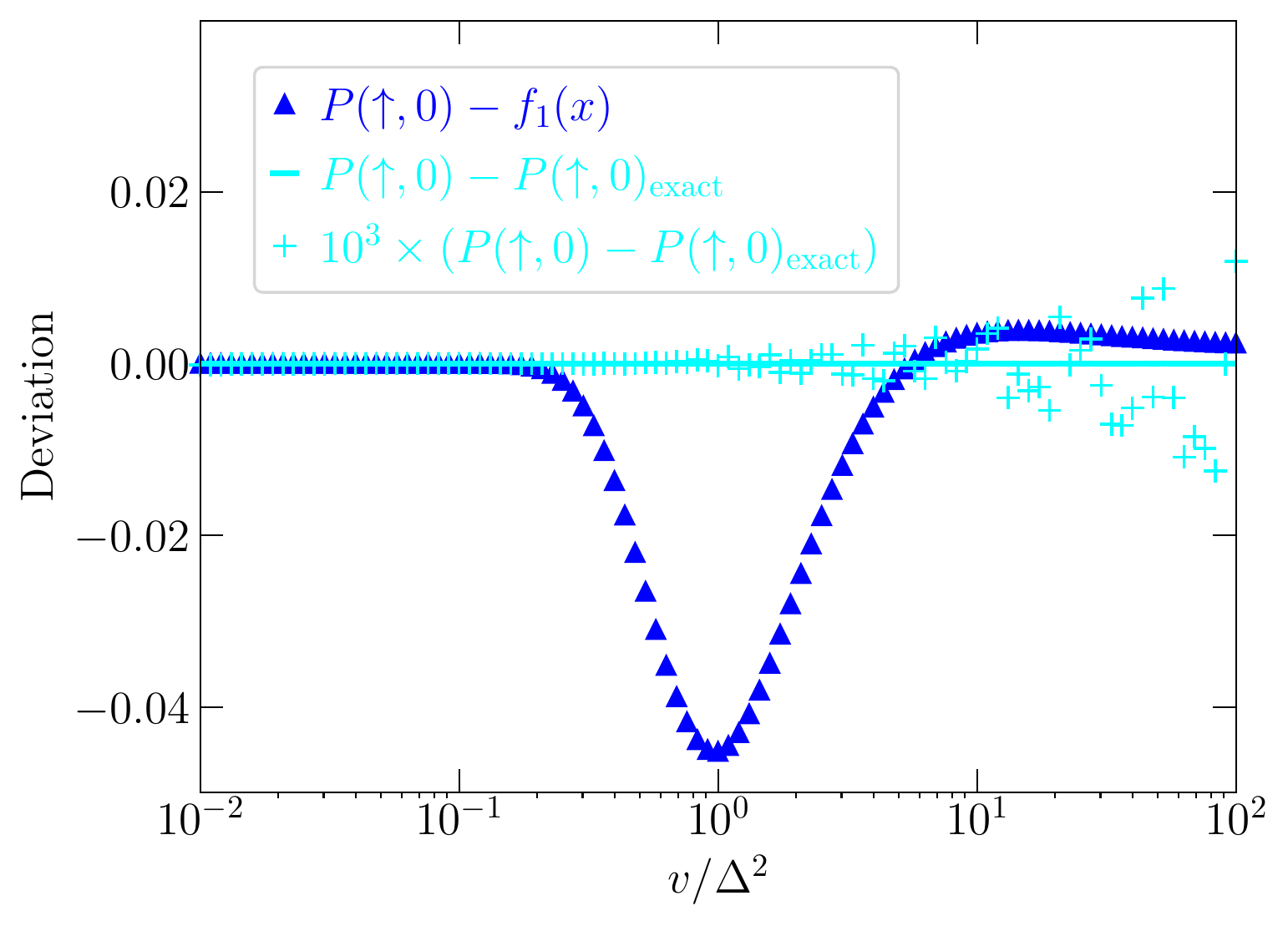}
\caption{Deviation between the data plotted in Fig.~\ref{Fig:LZwO010} and different theoretical lines. The blue triangles show the deviation from the function $f_1(x)$, i.e.~the green line in Fig.~\ref{Fig:LZwO010}. The horizontal cyan line, which essentially coincides with the $x$ axis, is the deviation between the data and the exact solution given by Eq.~(\ref{Eq:LZwOProbabilities}). The cyan + symbols are the same as the cyan line, but magnified by a factor of $10^3$. The comparison between the blue triangles and cyan symbols demonstrates a crucial point for our purposes: the deviation between the fitting function $f_1(x)$ and the data is more than 3 orders of magnitude larger than the numerical error level in the data. As a result, $f_1(x)$ cannot be the exact solution, even if it might be considered an acceptable or good fitting function for approximation purposes.}
\label{Fig:LZwO010Deviation}
\end{figure}

For $P(\uparrow, 0)$ the suggested solutions were
\begin{eqnarray}
f_1(x) & = & \exp(\cos(\exp(-1/x))) - 1.718281953032
\nonumber \\
f_2(x) & = & ( \cos(1.33333333333333 \times \exp(-1/x)) )^2
\nonumber \\
f_3(x) & = & ( \cos(\exp(-1/x)) )^2
\nonumber \\
f_4(x) & = & ( \cos(x \times \exp(-1/x)) )^2
\nonumber \\
f_5(x) & = & ( \cos(x^3) )^2
\nonumber \\
f_6(x) & = & 1
\nonumber \\
f_7(x) & = & 0.
\label{Eq:AIFSolutionsLZwO010}
\end{eqnarray}
The first of these solutions looks rather complicated, but it is in fact a rather good approximation for the data, as shown in Fig.~\ref{Fig:LZwO010}. Nevertheless, there is a clear deviation between the data and the fitting function around $v/\Delta^2=1$. This deviation is investigated more closely in Fig.~\ref{Fig:LZwO010Deviation}. Although $f_1(x)$ might be an acceptable approximation for the data, its deviation from the data is orders of magnitude larger than what can be attributed to numerical errors. The function $f_1(x)$ can therefore be disqualified as a candidate for the exact solution. This result illustrates the point made in Sec.~\ref{Sec:Method} about the fact that we can set the acceptable error level for the fitting function to the numerical error level in the computation that generates the data. If we had incorporated such a constraint into the SR algorithm, $f_1(x)$ would have been disqualified, because its small deviation from the data (which reaches a maximum value of $4.5\times 10^{-2}$ at $v/\Delta^2=1$) is clearly larger than the computational error level. The functions $f_2(x)$ and $f_3(x)$ are less good approximations, with clear deviations from the data for large values of $v/\Delta^2$. The functions $f_4(x)$ and $f_5(x)$ are close to 1 for small $v/\Delta^2$ values. When $P(\uparrow, 0)$ makes a turn and deviates away from 1, both $f_4(x)$ and $f_5(x)$ start to oscillate fast. To summarize, only the first suggested solution is a good approximation, but it is clearly not the exact solution that we are seeking. It is somewhat surprising that AIF did not find the exact solution given in Eq.~(\ref{Eq:LZwOProbabilities}), even though the exact solution is less complex than some of the solutions suggested by AIF.

\begin{figure}[h]
\includegraphics[width=10.0cm]{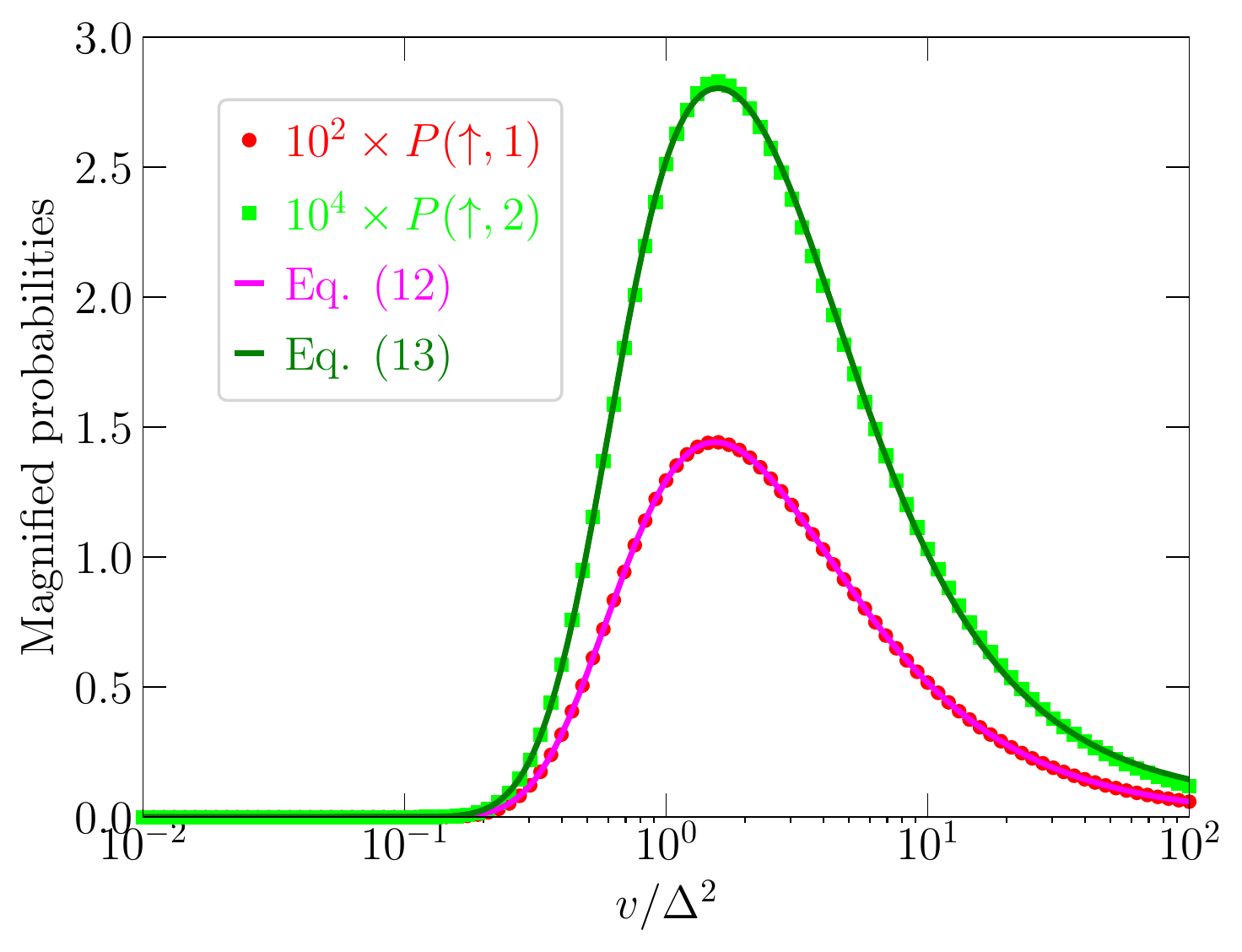}
\caption{Probabilities $P(\uparrow, 1)$ [red circles] and $P(\uparrow, 2)$ [green squares] as functions of $v/\Delta^2$ for the LZ problem of a TLS coupled to a harmonic oscillator with $g/\omega=0.1$. The magenta line is the best function generated by AIF to fit the $P(\uparrow, 1)$ data, and it is given by $f_1(x)$ in Eq.~(\ref{Eq:AIFSolutionsLZwO011}). The function $f_2(x)$, which is not plotted, is only slightly worse in fitting the data. The green line is the best function generated by AIF to fit the $P(\uparrow, 2)$ data, i.e.~the function $f(x)$ in Eq.~(\ref{Eq:AIFSolutionsLZwO012}). Although the fits look excellent, the deviations are larger than the level of computational errors, as is shown in Fig.~\ref{Fig:LZwO012Deviation}.}
\label{Fig:LZwO012}
\end{figure}

\begin{figure}[h]
\includegraphics[width=10.0cm]{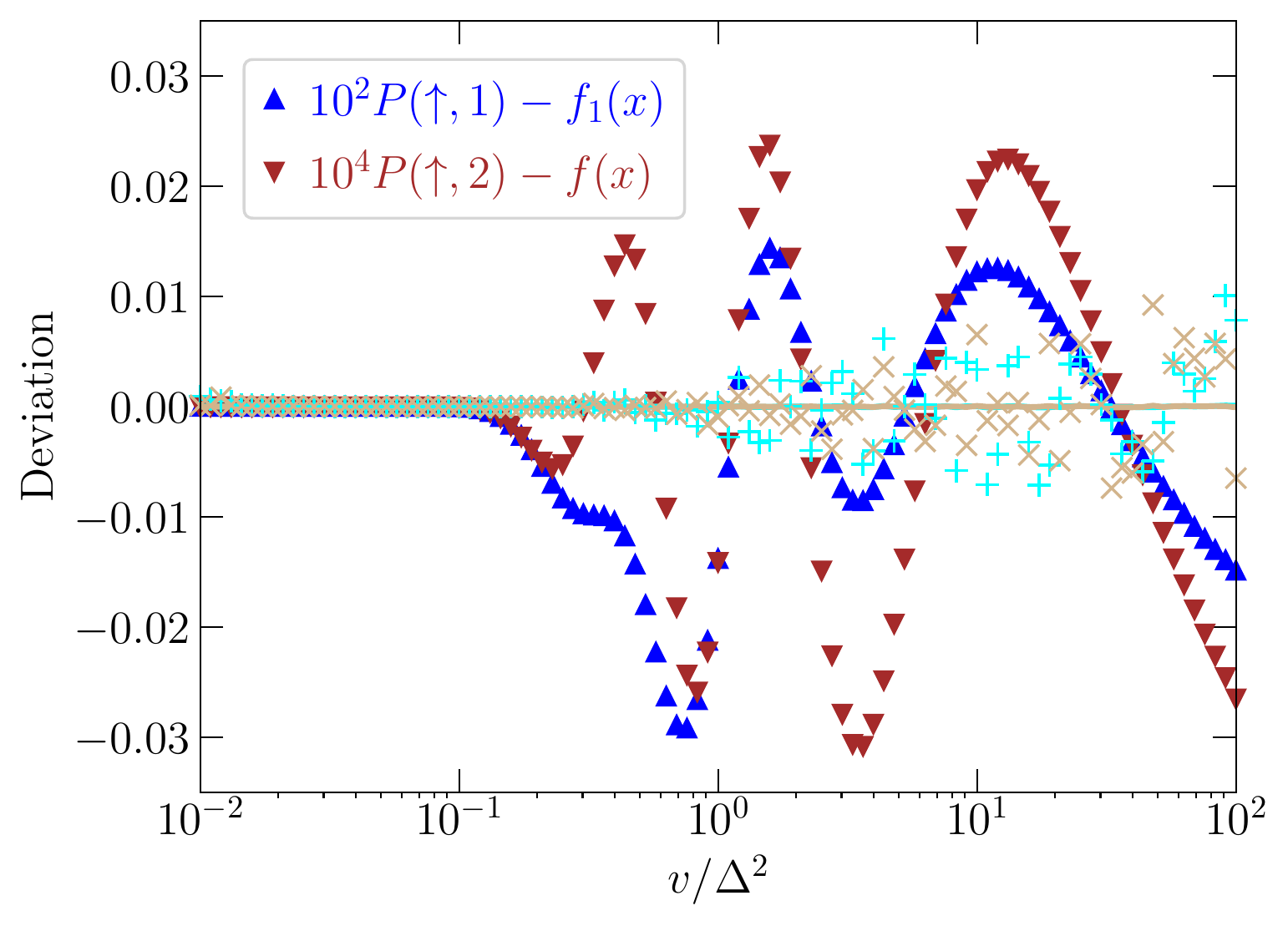}
\caption{Deviations between the data plotted in Fig.~\ref{Fig:LZwO012} and different theoretical lines. The blue and brown triangles show, respectively, the deviations between the red circles and green squares in Fig.~\ref{Fig:LZwO012} and the corresponding fitting functions. The cyan and tan lines, both of which barely deviate from the $x$ axis, are the deviations between the data and the exact solutions given by Eq.~(\ref{Eq:LZwOProbabilities}). The cyan + symbols and tan $\times$ symbols are the same as the cyan and tan lines, but magnified by a factor of $10^2$. The comparison between the different lines demonstrates that the deviations between the data and fitting functions in Fig.~\ref{Fig:LZwO012} are 2 orders of magnitude larger than the numerical error levels. In other words, although the fitting functions in Fig.~\ref{Fig:LZwO012} are good fitting functions, they cannot be the exact solutions that we are seeking.}
\label{Fig:LZwO012Deviation}
\end{figure}

For $P(\uparrow, 1)$ and $P(\uparrow, 2)$, after 10 runs for each data set, AIF gave only one suggested solution, namely 0 (expressed in different ways). Interestingly, on some of the runs on the $P(\uparrow, 2)$ data, the suggested solutions included variations on the expression
\begin{eqnarray}
f(x) = \sqrt{- \exp (-1/x^2)},
\label{Eq:AIFSolutionsLZwO012Imaginary}
\end{eqnarray}
which is clearly imaginary, although the input data for the fitting is all real. We suspected that the reason for the inability of AIF to generate any good fitting functions in these cases could be the fact that the probabilities are small, with maximum values of 0.0144 and 0.000283 for $P(\uparrow, 1)$ and $P(\uparrow, 2)$, respectively. We multiplied the probability data by $10^2$ and $10^4$, such that the peak value in each data set was amplified to the order of one. After this change, the best two suggested solutions for $P(\uparrow, 1)$ were
\begin{eqnarray}
f_1(x) & = & 1.69777143001556 \times \sin \left\{ \sin \left[ 3.12565946578979 \times \exp \left( \frac{-1.06572902202606}{x^{1.03745067119598}} \right) \right] \right\}
\nonumber \\
f_2(x) & = & 1.48647665977478 \times \sin \left\{ 3.11723256111145 \times \exp \left[ \frac{-1.05347275733948}{x} \right] \right\}.
\label{Eq:AIFSolutionsLZwO011}
\end{eqnarray}
For $P(\uparrow, 2)$, the best suggested solution was
\begin{eqnarray}
f(x) & = & 3.33333333333333 \times \sin \left\{ \sin \left[ 3.12629866600037 \times \exp \left( \frac{-1.10573124885559}{x^{1.04231333732605}} \right) \right] \right\}.
\nonumber \\
& &
\label{Eq:AIFSolutionsLZwO012}
\end{eqnarray}
As can be seen in Fig.~\ref{Fig:LZwO012}, Eqs.~(\ref{Eq:AIFSolutionsLZwO011}) and (\ref{Eq:AIFSolutionsLZwO012}) give very good fits to the data. However, although the deviations are barely discernible in Fig.~\ref{Fig:LZwO012}, closer inspection (Fig.~\ref{Fig:LZwO012Deviation}) reveals that the deviations are 2 orders of magnitude larger than the numerical error levels in the data. Hence both Eqs.~(\ref{Eq:AIFSolutionsLZwO011}) and (\ref{Eq:AIFSolutionsLZwO012}) can be disqualified from being the exact solutions that we are seeking.

\begin{figure}[h]
\includegraphics[width=10.0cm]{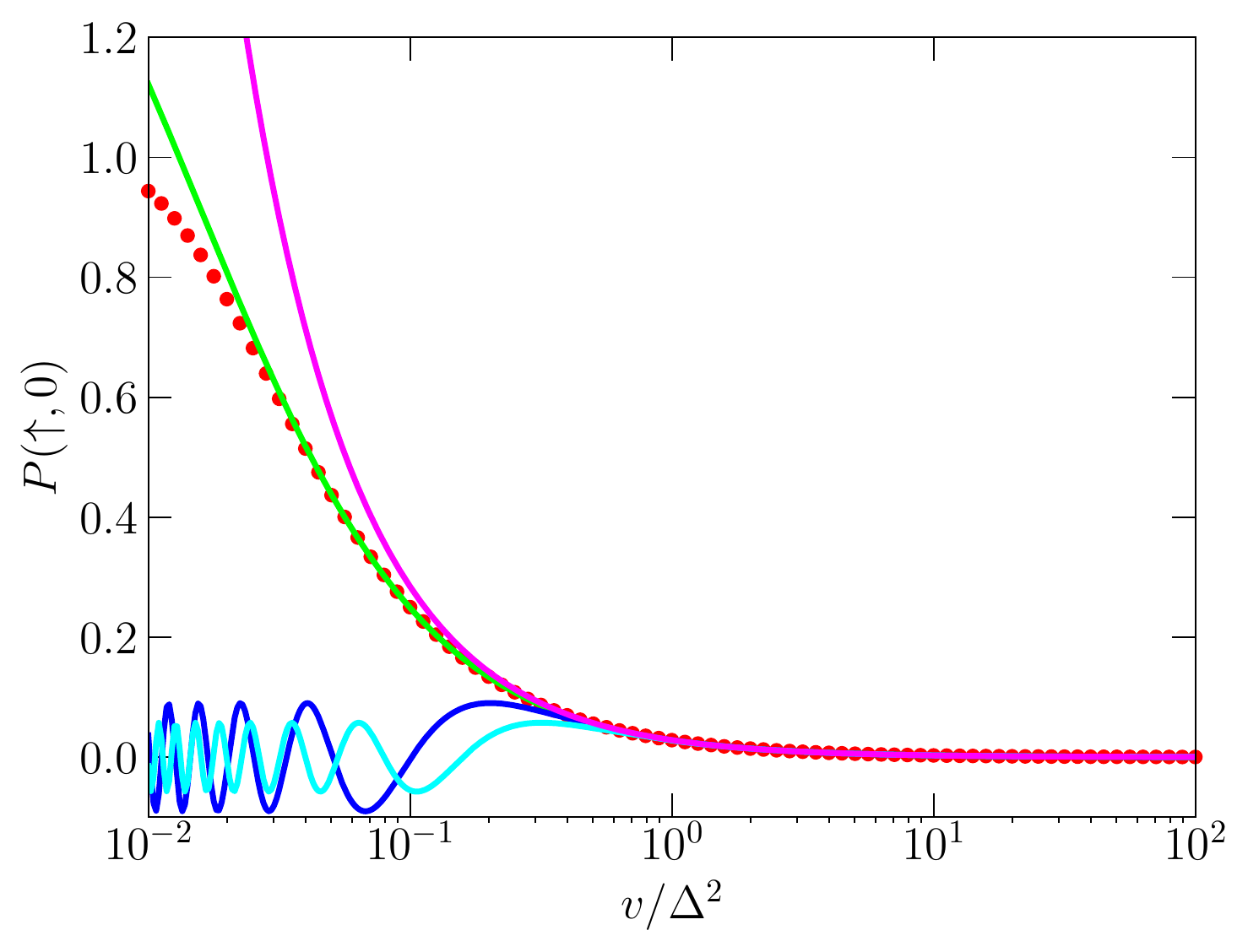}
\caption{Same as in Fig.~\ref{Fig:LZwO010}, but for $g/\omega=1$ and the fitting functions taken from Eq.~(\ref{Eq:AIFSolutionsLZwO100}). The best solution suggested by AIF, i.e.~the green line, is a good fit for large values of $v/\Delta^2$ but clearly fails for small values of $v/\Delta^2$. Furthermore, all the plotted solutions have values outside the physical range $0<P(\uparrow, 0)<1$ and hence cannot be the exact solution.}
\label{Fig:LZwO100}
\end{figure}

We performed the same calculations for $g/\omega=1$. The results for $P(\downarrow, 0)$ were similar to those shown in Eq.~(\ref{Eq:AIFSolutionsLZwO01E}): several equivalent expressions describing the correct solution (including zero or small extra terms) as well as a few low-complexity but incorrect solutions [specifically $f_4(x)$, $f_5(x)$ and $f_6(x)$]. For $P(\uparrow, 0)$ the suggested solutions were
\begin{eqnarray}
f_1(x) & = & 0.028778633465 \times \left( x+(\exp(\pi+1)+1)^{-1}\right)^{-1}
\nonumber \\
& \approx & \frac{0.028779}{x+0.015649}
\nonumber \\
f_2(x) & = & 0.090288630445 \times \sin \frac{1}{\pi x}
\nonumber \\
f_3(x) & = & 0.057509404466 \times \sin \frac{0.5}{x}
\nonumber \\
f_4(x) & = & \frac{0.028444255608}{x},
\label{Eq:AIFSolutionsLZwO100}
\end{eqnarray}
as well as a few additional solutions that have the same forms as some of those in Eq.~(\ref{Eq:AIFSolutionsLZwO100}) but with different constants. The function $f_1(x)$ is a good approximation for the data except for a small deviation at the small-$x$ end, as shown in Fig.~\ref{Fig:LZwO100}. It should be emphasized that this good agreement is somewhat misleading, as the data saturates at 1 in the limit $x\rightarrow 0$ while $f_1(x)$ continues to increase with decreasing $x$ until it reaches $f_1(0)=1.839$, a clearly unphysical and therefore disqualifying feature. This case illustrates two points that we mentioned in Sec.~\ref{Sec:Method}: (1) the usefulness of imposing physicality conditions on the allowed solutions and (2) the possibility of using active learning and adding more data points based on the initial results of fitting the first batch of data points. The functions $f_2(x)$ and $f_3(x)$ are good approximations for large values of $x$ (partly because both they and $P(\uparrow, 0)$ approach 0 in the limit $x\rightarrow\infty$) but become fast oscillating functions, with positive and negative values, at small values of $x$. Clearly this behavior is unphysical and disqualifying for these functions as candidates for the exact solution.

\begin{figure}[h]
\includegraphics[width=10.0cm]{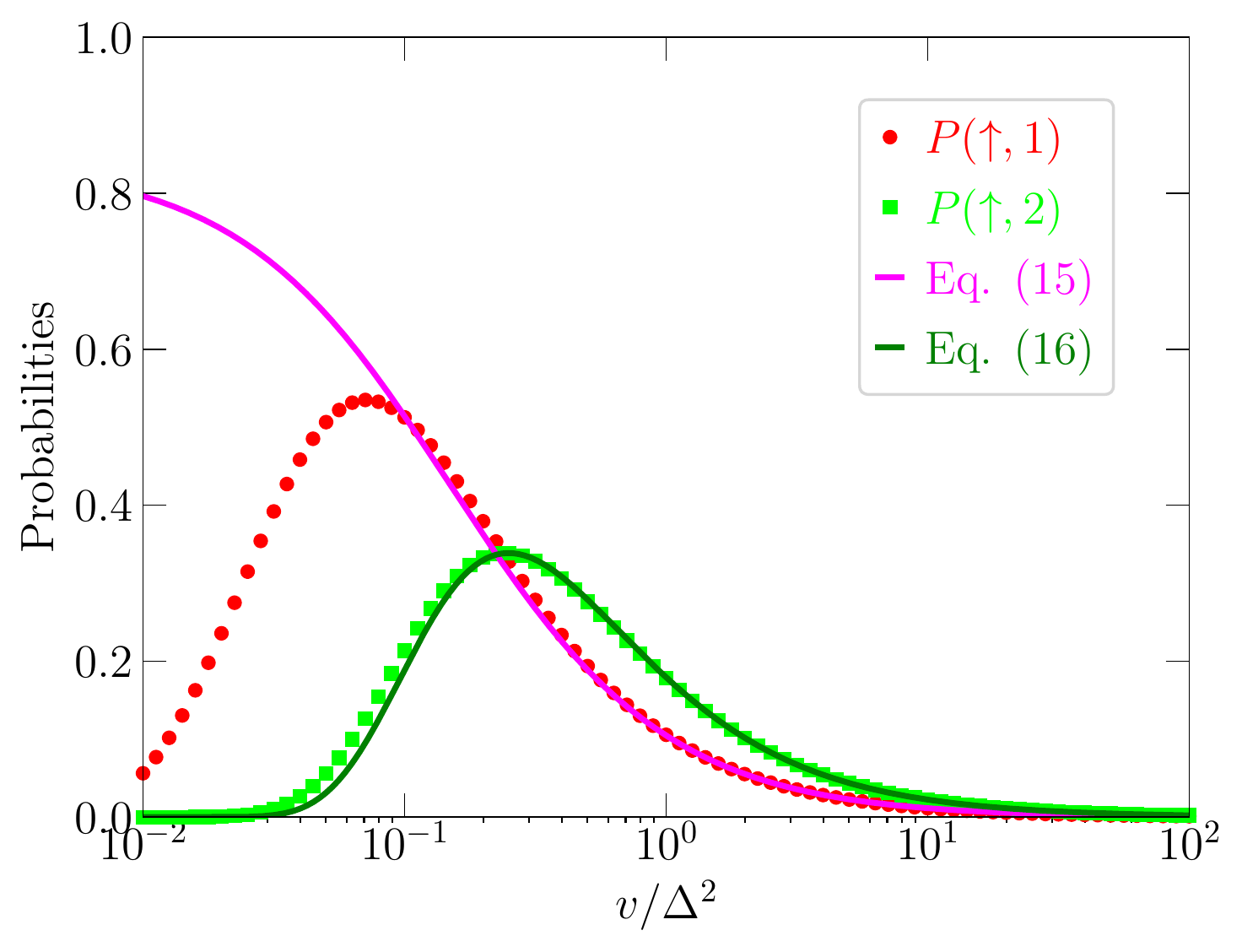}
\caption{Same as in Fig.~\ref{Fig:LZwO012}, but for $g/\omega=1$ and the fitting functions taken from Eqs.~(\ref{Eq:AIFSolutionsLZwO101}) and (\ref{Eq:AIFSolutionsLZwO102}). The fit of the $P(\uparrow, 1)$ data is poor, but the fit for the $P(\uparrow, 2)$ data is good. Nevertheless, there is a slight deviation between the data and the fitting function exceeding the numerical errors in the computation, meaning that the fitting function suggested by AIF is not the exact solution.}
\label{Fig:LZwO102}
\end{figure}

The data and best suggested solutions for $P(\uparrow, 1)$ and $P(\uparrow, 2)$ are shown in Fig.~\ref{Fig:LZwO102}. The best suggested solution to fit the $P(\uparrow, 1)$ data was
\begin{eqnarray}
f(x) = \sin \left( -0.000147037180 + \frac{1}{x \times \exp(\pi-1)+1} \right).
\label{Eq:AIFSolutionsLZwO101}
\end{eqnarray}
The best suggested solution for $P(\uparrow, 2)$ was
\begin{eqnarray}
f(x) = 0.230159951539 \times \frac{\exp(1/x)^{-0.25}}{x}.
\label{Eq:AIFSolutionsLZwO102}
\end{eqnarray}
The function in Eq.~(\ref{Eq:AIFSolutionsLZwO101}) is a poor fit to the data for small values of $x$. The function in Eq.~(\ref{Eq:AIFSolutionsLZwO102}) is a generally good fit to the data everywhere. However, the deviation between the data and the fitting function around $v/\Delta^2=10^{-1}$ is clearly sufficiently large to lead to the conclusion that the best suggested solution generated by AIF is not the exact solution that we are seeking.

\subsection{Unsolved problem -- Multi-level LZ problem}
\label{Sec:UnsolvedProblems}

One example of a problem that has no known analytic solution is the multi-level LZ problem, i.e.~the generalization of the LZ problem (described in Sec.~\ref{Sec:Introduction}) to a quantum system with more than two quantum states. For example, in the three-level case, the Schr\"{o}dinger equation can be expressed as:
\begin{equation}
\left( \begin{array}{c} i\frac{d\psi_1}{dt} \\ i\frac{d\psi_2}{dt} \\ i\frac{d\psi_3}{dt} \end{array} \right) =
\left( \begin{array}{ccc} v_1 t + E_1 & \Delta_{12}/2 & \Delta_{13}/2 \\ \Delta_{12}/2 & v_2 t + E_2 & \Delta_{23}/2 \\ \Delta_{13}/2 & \Delta_{23}/2 & v_3 t + E_3 \end{array} \right) \cdot \left( \begin{array}{c} \psi_1 \\ \psi_2 \\ \psi_3 \end{array} \right).
\label{Eq:ThreeLevelLZSE}
\end{equation}
This equation has nine parameters ($v_1, v_2, v_3, E_1, E_2, E_3, \Delta_{12}, \Delta_{13}$, and $\Delta_{23}$), but one can use simple arguments to reduce the number to five parameters that determine the solution \cite{MultiLevelLZFootnote}. The five parameters on which the unknown functions (i.e. $\left|\psi_j(t\rightarrow\infty)\right|^2$) will depend are $(v_3-v_+)/v_-, (E_3-E_+-E_-(v_3-v_+)/v_-)^2/v_-, \Delta_{12}^2/v_-, \Delta_{13}^2/v_-$, and $\Delta_{23}^2/v_-$, where $E_{\pm}=(E_1 \pm E_2)/2$ and $v_{\pm}=(v_1 \pm v_2)/2$. We can easily solve Eq.~(\ref{Eq:ThreeLevelLZSE}) numerically for any combination of parameters and obtain results that are accurate to several significant figures (although extreme parameter values that approach zero or infinity can pose problems for the numerical solution of the equation) \cite{Ashhab2016}.

It should be noted here that although the multi-level LZ problem is not fully solved, there is an exact analytic solution for the probability to remain in the same state if the system starts in the lowest or highest state. In other words, assuming that at $t\rightarrow -\infty$ the state is $(\psi_1, \psi_2, \psi_3) = (1,0,0)$ and $v_1$ is larger than both $v_2$ and $v_3$, we obtain the exact solution
\begin{equation}
|\psi_{1}(t\rightarrow\infty)|^2 = \exp \left\{-\sum_{j\neq 1}\frac{\pi\Delta_{1j}^2}{2(v_1 - v_j)} \right\},
\label{Eq:3LLZP1Exact}
\end{equation}
where the sum over $j$ is a sum over all the states except the initial state \cite{Shytov}.

Since we expect that this problem in its full five-variable form is too hard for the current version of AIF, we consider just one special case: we set $\Delta_{12}=\Delta_{13}=\Delta_{23}=\Delta$, $v_1=-v_3=v$, $v_2=0$, $E_1=E_3$, $E_2-E_1=-\Delta/2$. As explained in Sec.~\ref{Sec:Introduction}, the final occupation probabilities will now be functions of the single parameter $v/\Delta^2$. These are the probabilities that we would like to evaluate and subsequently fit. We emphasize that even this special case of the three-level LZ problem has no known analytical solution.

Similarly to what we did with the TLS-oscillator problem treated in Sec.~\ref{Sec:SolvedProblems}, we generate the data points by numerical integration of the Schr\"odinger equation, which is Eq.~(\ref{Eq:ThreeLevelLZSE}) in this case. For the numerical calculations, we set the initial and final times to $-vt_{\rm initial}/\Delta=vt_{\rm final}/\Delta=10^5$. We divide the total time into $10^7$ time steps and use the average Hamiltonian in each step. We assume that the initial state is $(\psi_1, \psi_2, \psi_3) = (1,0,0)$ and calculate the probabilities $P_1$, $P_2$ and $P_3$ ($P_n=|\psi_n|^2$) at the final time. Generating a single data point takes a few minutes, which does not create a computational bottleneck for our overall calculation. For comparison, we repeated the calculations by setting $-vt_{\rm initial}/\Delta=vt_{\rm final}/\Delta=10^4$ and using $10^5$ time steps. With these computation settings, generating one data point takes a few seconds. By comparing the different simulation results with the formula in Eq.~(\ref{Eq:3LLZP1Exact}), we can estimate that our final probability data obtained with the finer simulation parameters deviate from the exact values by up to $4\times 10^{-4}$, while those with the coarser simulation parameters deviate from the exact values by up to $9\times 10^{-3}$ (mainly at the small-$v/\Delta^2$ end of the data in both cases).

\begin{figure}[h]
\includegraphics[width=10.0cm]{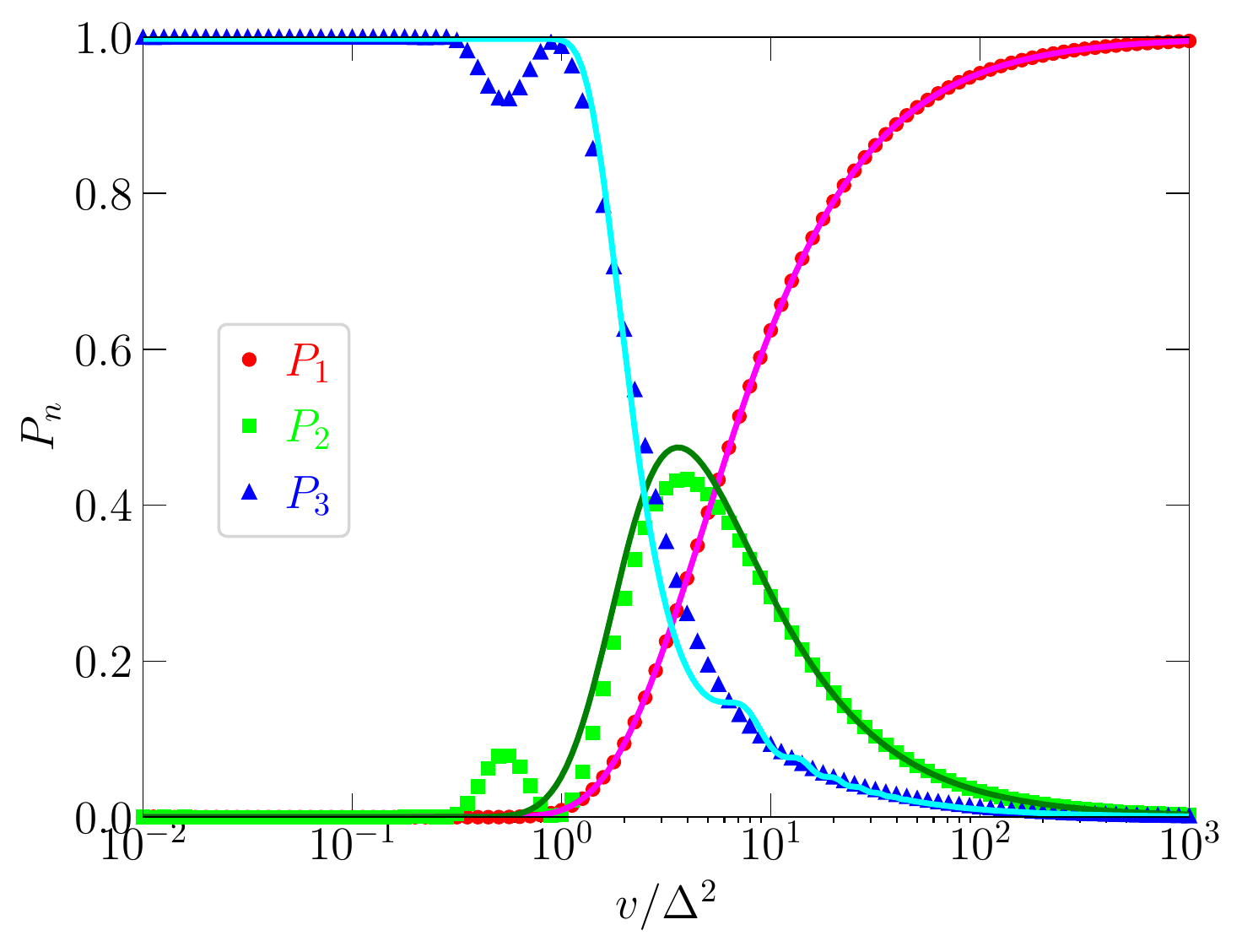}
\caption{Probabilities $P_n$ as functions of $v/\Delta^2$ for the three-level LZ problem described in Eq.~(\ref{Eq:ThreeLevelLZSE}) with $\Delta_{12}=\Delta_{13}=\Delta_{23}=\Delta$, $v_1=-v_3=v$, $v_2=0$, $E_1=E_3$, $E_2-E_1=-\Delta/2$. The red circles, green squares and blue triangles are, respectively, the $P_1$, $P_2$ and $P_3$ data points that were generated by numerically solving the Schr\"odinger equation. The $P_1$ data is very well fitted by the magenta line, which is given by Eq.~(\ref{Eq:3LLZP1Fit}). The $P_2$ and $P_3$ data are generally well fitted by the green and cyan lines, respectively. These lines are given, respectively, by Eqs.~(\ref{Eq:3LLZP2Fit}) and (\ref{Eq:3LLZP3Fit}). In spite of the overall agreement, there are large deviations in the cases of both $P_2$ and $P_3$, especially in the region where the data points have a peak-dip pair in the region $0.3<v/\Delta^2<1$, meaning that Eqs.~(\ref{Eq:3LLZP2Fit}) and (\ref{Eq:3LLZP3Fit}) cannot be the exact solutions for $P_2$ and $P_3$.}
\label{Fig:LZ3L}
\end{figure}

The data and best suggested solutions are shown in Fig.~\ref{Fig:LZ3L}. For the $P_1$ data set, the best suggested solution by AIF was
\begin{equation}
f(x) = \exp \left\{ -4.71008510610149 \times x^{-0.999879062175751} \right\}.
\label{Eq:3LLZP1Fit}
\end{equation}
This function deviates from the data by $4\times 10^{-4}$, which is the numerical error level. In other words the function in Eq.~(\ref{Eq:3LLZP1Fit}) meets the main criterion that the exact solution must satisfy. Indeed, we can recognize that the constants in Eq.~(\ref{Eq:3LLZP1Fit}) are close to simple constants (to within $5\times 10^{-4}$), and the formula can be simplified by rounding the constants to
\begin{equation}
f(x) = \exp \left\{ -\frac{3\pi}{2x}\right\}.
\end{equation}
This function coincides with the known solution for $P_1$. For $P_2$ the best solution suggested by AIF was
\begin{equation}
f(x) = \sin \frac{3.340054886514}{x + \exp((\pi+1)/x)}.
\label{Eq:3LLZP2Fit}
\end{equation}
For $P_3$ the three top suggestions by AIF were all variants of the function
\begin{equation}
f(x) = 1 - \exp \left\{ \frac{1}{-x + \sin x} \right\},
\label{Eq:3LLZP3Fit}
\end{equation}
with some negligibly small differences in the constants. The functions in Eqs.~(\ref{Eq:3LLZP2Fit}) and (\ref{Eq:3LLZP3Fit}) are good fits for the data for most values of $v/\Delta^2$. However, these functions completely miss the peak-dip pair in the data in the region $0.3<v/\Delta^2<1$ and exhibit a noticeable deviation from the data in the region $2<v/\Delta^2<10$ as well. Even without any statistical analysis of the data-fit deviations, the fact that the fitting functions miss the peak-dip feature is sufficient for us to make a statement about these functions. Peaks in physical quantities always represent physical phenomena. If a fitting function misses such a feature, the function is missing a physical effect that occurs in the system under study. On the other hand, the fitting function for $P_3$, i.e.~Eq.~(\ref{Eq:3LLZP3Fit}), exhibits oscillatory behavior that is most clearly visible around $v/\Delta^2\sim 10$, even though the data does not seem to have any noticeable oscillations in that region. Hence the fitting functions in Eqs.~(\ref{Eq:3LLZP2Fit}) and (\ref{Eq:3LLZP3Fit}) cannot be the exact solutions that we are seeking.

To conclude this section, we emphasize that the examples that we have presented here are just a few specific examples that are simple generalizations of the LZ problem. There will undoubtedly be more important and interesting problems in theoretical physics to tackle using SR when the power of the algorithms increases and can identify more complex functions. Any unsolved problem for which we can use computational methods to find the solution for any set of input variables will be well suited for treating with this approach.

\section{Conclusion}
\label{Sec:Conclusion}

We have explored the use of SR tools to find exact closed-form solutions for analytically posed problems in theoretical physics and applied mathematics. Assuming that a problem can be solved numerically for any set of input variables, the approach is to first generate a numerical data set from the analytically posed problem and then use SR, possibly combined with active learning, to identify the functional representation of the data, hence obtaining the analytic solution to the original problem. We gave a few examples based on the LZ problem to illustrate the computational procedure and demonstrate the abilities and limitations of state-of-the-art SR tools. Our results suggest that modifying SR packages to apply stricter constraints on the allowed error and to apply physicality constraints will be crucial to make these packages more capable and efficient in finding exact solutions. Developing new algorithms in which these modifications are incorporated is beyond the scope of the present work. We hope that our results will motivate researchers to take our observations into consideration when developing future SR algorithms. Although our focus was on finding exact solutions, several of the implementation ideas that we discussed in this work can also be utilized in the search for approximations that incorporate physical constraints. We believe that the ideas discussed here will help in the development of powerful computational tools to help answer important questions in theoretical physics and applied mathematics.

\section*{Acknowledgment}

We would like to thank Sanjay Chawla, Lior Horesh, Raka Jovanovic, Stefano Rizzo and Kouichi Semba for useful discussions and Takashi Nakayama for helping set up the computing environment that was used for some of our calculations. This work was partially supported by the Q-LEAP Quantum AI Flagship Project of the Ministry of Education, Culture, Sports, Science and Technology (MEXT), Japan (Grant Number JPMXS0120319794).

\section*{Data availability}

The datasets generated and/or analyzed during the current study are available from the author on reasonable request.


\begin{thebibliography}{99}

\bibitem{Silver} D. Silver, T. Hubert, J. Schrittwieser, I. Antonoglou, M. Lai, A. Guez, M. Lanctot, L. Sifre, D. Kumaran, T. Graepel, T. Lillicrap, K. Simonyan, and D. Hassabis, A general reinforcement learning algorithm that masters chess, shogi, and Go through self-play, Science {\bf 362}, 1140 (2018).

\bibitem{NatMatEditorial} Ascent of machine learning in medicine, Nat. Mater. {\bf 18}, 407 (2019)

\bibitem{Radovic} A. Radovic, M. Williams, D. Rousseau, M. Kagan, D. Bonacorsi, A. Himmel, A. Aurisano, K. Terao, and T. Wongjirad, Machine learning at the energy and intensity frontiers of particle physics, Nature {\bf 560}, 41 (2018).

\bibitem{Carleo} G. Carleo, I. Cirac, K. Cranmer, L. Daudet, M. Schuld, N. Tishby, L. Vogt-Maranto, and L. Zdeborov\'a, Machine learning and the physical sciences, Rev. Mod. Phys. {\bf 91}, 045002 (2019).

\bibitem{Iten} R. Iten, T. Metger, H. Wilming, L. del Rio, and R. Renner, Discovering physical concepts with neural networks, Phys. Rev. Lett. {\bf 124}, 010508 (2020).

\bibitem{Qiu} J. Qiu, G. Zhong, Y. Lu, K. Xin, H. Qian, and X. Zhu, The Newton scheme for deep learning, arXiv:1810.07550.

\bibitem{Alhousseini} I. Alhousseini, W. Chemissany, F. Kleit, and A. Nasrallah, Physicist's journeys through the AI world - a topical review. There is no royal road to unsupervised learning, arXiv:1905.01023.

\bibitem{Koza} J. R. Koza, {\it Genetic Programming: On the Programming of Computers by Means of Natural Selection}, (MIT Press, Cambridge, 1992).

\bibitem{Schoenauer} M. Schoenauer, M. Sebag, F. Jouve, B. Lamy, and H. Maitournam, Evolutionary identification of macro-mechanical models, in {\it Advances in Genetic Programming}, P. J. Angeline and K. E. Kinnear Eds. (MIT Press, 1996).

\bibitem{Schmidt} M. Schmidt and H. Lipson, Distilling free-form natural laws from experimental data, Science {\bf 324}, 81 (2009).

\bibitem{Eureqa} R. Dub\v{c}\'{a}kov\'{a}, Genetic programming and evolvable machines {\bf 12}, 173 (2011).

\bibitem{Gaucel} S. Gaucel, M. Keijzer, E. Lutton, and A. P. Tonda, Learning dynamical systems using standard symbolic regression, EuroGP 25 (2014).

\bibitem{Wu} T. Wu and M. Tegmark, Toward an artificial intelligence physicist for unsupervised learning, Phys. Rev. E {\bf 100}, 033311 (2019).

\bibitem{Jovanovic} R. Jovanovic and S. Ashhab, A GRASP approach  for symbolic regression, Proceedings of the 2019 IEEE Symposium Series on Computational Intelligence (SSCI), 1723 (2019).

\bibitem{Hernandez} A. Hernandez, A. Balasubramanian, F. Yuan, S. Mason, and T. Mueller, Fast, accurate, and transferable many-body interatomic potentials by symbolic regression, npj Comput. Mater. {\bf  5}, 112 (2019).

\bibitem{Guimera} R. Guimerà, I. Reichardt, A. Aguilar-Mogas, F. A. Massucci, M. Miranda, J. Pallarès, and M. Sales-Pardo, A Bayesian machine scientist to aid in the solution of challenging scientific problems, Sci. Adv. {\bf 6}, eaav6971 (2020).

\bibitem{Udrescu} S.-M. Udrescu and M. Tegmark, AI Feynman: A physics-inspired method for symbolic regression, Sci. Adv. {\bf 6}, eaay2631 (2020).

\bibitem{Udrescu2} S.-M. Udrescu, A. Tan, J. Feng, O. Neto, T. Wu, and M. Tegmark, AI Feynman 2.0: Pareto-optimal symbolic regression exploiting graph modularity, arXiv:2006.10782 [34th Conference on Neural Information Processing Systems (Neurips 2020), Vancouver, Canada].

\bibitem{Bomarito} G. F. Bomarito, T. S. Townsend, K. M. Stewart, K. V. Esham, J. M. Emery, and J. D. Hochhalter, Development of interpretable, data-driven plasticity models with symbolic regression, Comput. Struct. {\bf 252}, 106557 (2021).

\bibitem{Keren} L. S. Keren, A. Liberzon, and T. Lazebnik, A computational framework for physics-informed symbolic regression with straightforward integration of domain knowledge, Sci. Rep. {\bf 13}, 1249 (2023).

\bibitem{Fajardo} O. Fajardo-Fontiveros, I. Reichardt, H. R. De Los Ríos, J. Duch, M. Sales-Pardo, and R. Guimerà, Fundamental limits to learning closed-form mathematical models from data, Nature Commun. {\bf 14}, 1043 (2023).

\bibitem{Cornelio} C. Cornelio, S. Dash, V. Austel, T. R. Josephson, J. Goncalves, K. L. Clarkson, N. Megiddo, B. El Khadir, and L. Horesh, Combining data and theory for derivable scientific discovery with AI-Descartes, Nature Commun. {\bf 14}, 1777 (2023).

\bibitem{Makke} N. Makke and S. Chawla, Interpretable scientific discovery with symbolic regression: a review, Artif. Intell. Rev. {\bf 57}, 2 (2024).

\bibitem{Kubalik2023} J. Kubal\'{i}k, E. Derner, and R. Babu\v{s}ka, Toward physically plausible data-driven models: a novel neural network approach to symbolic regression, IEEE Access {\bf 11}, 61481 (2023).

\bibitem{Oh} H. Oh, R. Amici, G. Bomarito, S. Zhe, R. M. Kirby, and J. Hochhalter, Inherently interpretable machine learning solutions to diﬀerential equations, arXiv:2302.03175.

\bibitem{Cao} L. Cao, Z. Zheng, C. Ding, J. Cai, and M. Jiang, Genetic programming symbolic regression with simplification-pruning operator for solving differential equations, ICONIP (8) 287 (2023).

\bibitem{CoryWright} R. Cory-Wright, B. El Khadir, C. Cornelio, S. Dash, and L. Horesh, AI Hilbert: a new paradigm for scientific discovery by unifying data and background knowledge, arXiv:2308.09474.

\bibitem{DiGiacomo} F. Di Giacomo and E. Nikitin, The Majorana formula and the Landau–Zener–St\"uckelberg treatment of the avoided crossing problem, Sov. Phys. Uspekhi {\bf 48}, 515 (2005).

\bibitem{Shevchenko} S. N. Shevchenko, S. Ashhab, and F. Nori, Landau-Zener-St\"uckelberg interferometry, Phys. Rep. {\bf 492}, 1 (2010).

\bibitem{Ashhab2010} S. Ashhab and F. Nori, Qubit-oscillator systems in the ultrastrong-coupling regime and their potential for preparing nonclassical states, Phys. Rev. A {\bf 81}, 042311 (2010).

\bibitem{Braak} D. Braak, Integrability of the Rabi model, Phys. Rev. Lett. {\bf 107}, 100401 (2011).

\bibitem{Unke} O. T. Unke, S. Chmiela, H. E. Sauceda, M. Gastegger, I. Poltavsky, K. T. Sch\"utt, A. Tkatchenko, and K.-R. M\"uller, Machine learning force fields, Chem. Rev. {\bf 121}, 10142 (2021).

\bibitem{Mishin} Y. Mishin, Machine-learning interatomic potentials for materials science, Acta Mater. {\bf 214}, 116980 (2021).

\bibitem{Behler} J. Behler, Four generations of high-dimensional neural network potentials, Chem. Rev. {\bf 121}, 10037 (2021).

\bibitem{Wolfram} S. Wolfram, {\it A New Kind of Science}, (Wolfram Media, Champaign, Illinois, 2002).

\bibitem{Rowland} T. Rowland, Computational irreducibility, from MathWorld--A Wolfram Web Resource, created by E. W. Weisstein, https://mathworld.wolfram.com/ComputationalIrreducibility.html

\bibitem{Settles} B. Settles, Computer Sciences Technical Report 1648 (2010); http://burrsettles.com/pub/settles.activelearning.pdf

\bibitem{Wubs} M. Wubs, K. Saito, S. Kohler, P. H\"anggi, and Y. Kayanuma, Gauging a quantum heat bath with dissipative Landau-Zener transitions, Phys. Rev. Lett. {\bf 97}, 200404 (2006).

\bibitem{Demkov} Yu. N. Demkov and V. I. Osherov, Stationary and nonstationary problems in quantum mechanics that can be solved by means of contour integration, Sov. Phys. JETP {\bf 53}, 1589 (1968).

\bibitem{Ashhab2023} S. Ashhab, T. Fuse, F. Yoshihara, S. Kim, and K. Semba, Controlling qubit-oscillator systems using linear parameter sweeps, New J. Phys. {\bf 25}, 093011 (2023).

\bibitem{Ashhab2016} S. Ashhab, Landau-Zener transitions in an open multilevel quantum system, Phys. Rev. A {\bf 94}, 042109 (2016).

\bibitem{Shytov} A. V. Shytov, Landau-Zener transitions in a multilevel system: An exact result, Phys. Rev. A {\bf 70}, 052708 (2004).

\bibitem{MultiLevelLZFootnote} If we do not consider the phases in the quantum superpositions at the initial and final times, we now have four combinations of initial and final states. We can for example take the initial state at $t\rightarrow -\infty$ as $(\psi_1, \psi_2, \psi_3) = (1,0,0)$ or $(\psi_1, \psi_2, \psi_3) = (0,1,0)$. In each case we have two unknown functions, e.g.~the values of $|\psi_2|^2$ and $|\psi_3|^2$ at $t\rightarrow \infty$. The remaining five probabilities can then be determined from unitarity and probability conservation, e.g.~the constraint that $|\psi_1|^2+|\psi_2|^2+|\psi_3|^2=1$.

\end{thebibliography}
\end{document}